\documentclass[12pt]{article}
\usepackage{times}
\usepackage{geometry}
\usepackage[utf8]{inputenc}
\usepackage{enumitem,amssymb}
\usepackage{ragged2e}
\newlist{thematic}{itemize}{8}
\setlist[thematic]{label=$\square$}
\usepackage{pifont}
\usepackage{amsmath}
\usepackage{hyperref}
\usepackage{wrapfig}
\usepackage{graphicx}
\usepackage{upgreek}
\usepackage{textcomp}
\usepackage{lineno}
\usepackage{color}
\usepackage[sort&compress,numbers]{natbib}
\usepackage{soul}
\usepackage{authblk}

\usepackage{lipsum}

\begin{document}
\begin{raggedright}
\Large
Astro2020 Science White Paper \linebreak

{\bf Cosmic-ray Antinuclei as Messengers for Dark Matter} \linebreak
\normalsize

\noindent \textbf{Thematic Areas:} \\
Cosmology and Fundamental Physics, Multi-Messenger Astronomy and Astrophysics\\
\vspace{0.2in}
  
\textbf{Authors:}\\
\vspace{0.1in}

Kerstin Perez$^1$,  Philip von Doetinchem$^2$, Tsuguo Aramaki$^3$, Mirko Boezio$^4$, Steven E. Boggs$^5$, William W. Craig$^6$, Lorenzo Fabris$^7$, Hideyuki Fuke$^8$, Florian Gahbauer$^6$, Charles J. Hailey$^6$, Rene Ong$^9$ 

\vspace{0.1in}
\textsuperscript{1}\,Massachusetts Institute of Technology (kmperez@mit.edu)\\
\textsuperscript{2}\,University of Hawaii at Manoa (philipvd@hawaii.edu)\\
\textsuperscript{3}\, Stanford Linear Accelerator Center\\
\textsuperscript{4}\, INFN, Sezione di Trieste\\
\textsuperscript{5}\, University of California San Diego\\
\textsuperscript{6}\, Columbia University\\
\textsuperscript{7}\, Oak Ridge National Laboratory\\
\textsuperscript{8}\, Institute of Space and Astronautical Science, Japan Aerospace Exploration Agency\\
\textsuperscript{9}\, University of California Los Angeles\\

\vspace{0.2in}

\textbf{Abstract:}

The origin of dark matter is a driving question of modern physics. Low-energy antideuterons provide a ``smoking gun" signature of dark matter annihilation or decay, essentially free of astrophysical background. Low-energy antiprotons are a vital partner for this analysis, and low-energy antihelium could provide further discovery space for new physics. In the coming decade, AMS-02 will continue accumulating the large statistics and systematic understanding necessary for it to probe rare antinuclei signatures, and GAPS, which is the first experiment optimized specifically for low-energy cosmic antinuclei, will begin several Antarctic balloon campaigns. The connection of cosmic-ray antinuclei and dark matter is reviewed and the outlook in light of experimental progress is presented.

\vspace{0.2in}

\textbf{Endorsers:}\\
\vspace{0.1in}
Steve Ahlen$^{10}$, 
Steve Barwick$^{11}$, 
Ralph Bird$^{9}$, 
Mathieu Boudaud$^{12,13}$, 
Marco Cirelli$^{12,13}$, 
Adam Coogan$^{14}$, 
Mingyang Cui$^{15}$, 
Amaresh Datta$^{2}$, 
Fiorenza Donato$^{16}$,
Carmelo Evoli$^{17}$, 
Laura Fabbietti$^{18}$, 
Nicolao Fornengo$^{16,19}$, 
Marek Gazdzicki$^{20}$, 
Diego Mauricio Gomez Coral$^{21}$, 
Dan Hooper$^{22,23}$, 
Alessandro Ibarra$^{17}$, 
Michael Kachelriess$^{24}$, 
Alexander Kalweit$^{25}$, 
Michael Korsmeier$^{26,27}$, 
Masayoshi Kozai$^{8}$, 
Sujie Lin$^{28}$, 
David Maurin$^{29}$, 
Arturo Menchaca-Rocha$^{21}$, 
Samuel Adam Isaac Mognet$^{30}$, 
Igor V. Moskalenko$^{31}$, 
Li Nan$^{32}$, 
Kenny C. Y. Ng$^{33}$, 
Alberto Oliva$^{34}$, 
Sergey Ostapchenko$^{35}$, 
Tanguy Pierog$^{36}$, 
Vivian Poulin$^{37,38}$, 
Stefano Profumo$^{39}$, 
Sean Quinn$^{9}$, 
Are Raklev$^{40}$, 
Marco Regis$^{18}$, 
Alberto Ribon$^{25}$, 
Pierre Salati$^{41}$, 
Pasquale D. Serpico$^{41}$, 
Achim Stoessl$^{2}$, 
Greg Tarle$^{42}$, 
Tim M.P. Tait$^{11}$, 
Nicola Tomassetti$^{19,43}$, 
Pierro Ullio$^{44}$, 
Michael Unger$^{36}$, 
Alfredo Urbano$^{19}$, 
Vladimir Uzhinsky$^{25,45}$, 
Manuela Vecchi$^{46,47}$, 
Andrea Vittino$^{27}$, 
Christoph Weniger$^{14}$, 
Scott Wakely$^{23}$, 
Martin Winkler$^{48}$, 
Dennis Wright$^{3}$, 
Mengjiao Xiao$^{1}$, 
Tetsuya Yoshida$^{8}$ 

\vspace{0.1in}
\textsuperscript{10}\,Boston University, 
\textsuperscript{11}\,University of California Irvine, 
\textsuperscript{12}\,Laboratoire de Physique Th\'eorique et Hautes Energies, 
\textsuperscript{13}\,CNRS-Sorbonne Universit\'e,
\textsuperscript{14}\,University of Amsterdam, 
\textsuperscript{15}\,Purple Mountain Observatory, 
\textsuperscript{16}\,University of Torino, 
\textsuperscript{17}\,Gran Sasso Science Institute, 
\textsuperscript{18}\,Technische Universit\"at M\"unchen, 
\textsuperscript{19}\, Istituto Nazionale di Fisica Nucleare, 
\textsuperscript{20}\,Goethe-University Frankfurt am Main, Jan Kochanowski University Kielce, 
\textsuperscript{21}\,Universidad Nacional Aut\'onoma de M\'exico, 
\textsuperscript{22}\,Fermilab, 
\textsuperscript{23}\,University of Chicago, 
\textsuperscript{24}\,Norges teknisk-naturvitenskaplige universitet , 
\textsuperscript{25}\,CERN, 
\textsuperscript{26}\,University of Turin, 
\textsuperscript{27}\,RWTH Aachen University,
\textsuperscript{28}\,Institute of High Energy Physics, 
\textsuperscript{29}\,Laboratoire de Physique Subatomique et Cosmologie,
\textsuperscript{30}\,Pennsylvania State University, 
\textsuperscript{31}\,Stanford University,
\textsuperscript{32}\,Chinese Academy of Sciences, 
\textsuperscript{33}\,Weizmann Institute of Science
\textsuperscript{34}\,Centro de Investigaciones Energ\'eticas, Medioambientales y Tecnol\'ogicas
\textsuperscript{35}\,Goethe-University, 
\textsuperscript{36}\,Karlsruhe Institute of Technology, 
\textsuperscript{37}\,Universit\'e de Savoie, 
\textsuperscript{38}\,Johns Hopkins University,
\textsuperscript{39}\,University of California Santa Cruz,
\textsuperscript{40}\,University of Oslo, 
\textsuperscript{41}\,Laboratoire d'Annecy-le-Vieux de Physique Th\'eorique, 
\textsuperscript{42}\,University of Michigan, 
\textsuperscript{43}\,Perugia University,
\textsuperscript{44}\,Scuola Internazionale Superiore di Studi Avanzati, 
\textsuperscript{45}\,Joint Institute for Nuclear Research,
\textsuperscript{46}\,University of Groningen,
\textsuperscript{47}\,University of S\"ao Paulo,
\textsuperscript{48}\,Stockholm University

\end{raggedright}

\pagebreak
\section{Cosmic-ray Antinuclei and Dark Matter}
\vspace{-0.1in}
\subsection{Introduction}

The origin of dark matter (DM) is a driving question of modern physics. Low-energy antideuterons provide a ``smoking gun" signature of DM annihilation or decay, essentially free of astrophysical background. A first-time detection of low-energy cosmic antideuterons would be an unambiguous signal of new physics, opening a transformative new field of cosmic-ray research and probing a variety of DM models that evade or complement collider, direct, or other cosmic-ray searches. Low-energy antiprotons are a vital partner for this analysis, and low-energy antihelium could provide further discovery space for new physics. In an era with no discovery of supersymmetry at the LHC or Weakly interacting massive particles (WIMPs) in ton-scale direct detection experiments, antinuclei signatures of DM have maintained their relevance in a wide variety of theories.

\begin{wrapfigure}{r}{0.52\linewidth}
\vspace{-0.1in}
\includegraphics[width=1\linewidth]{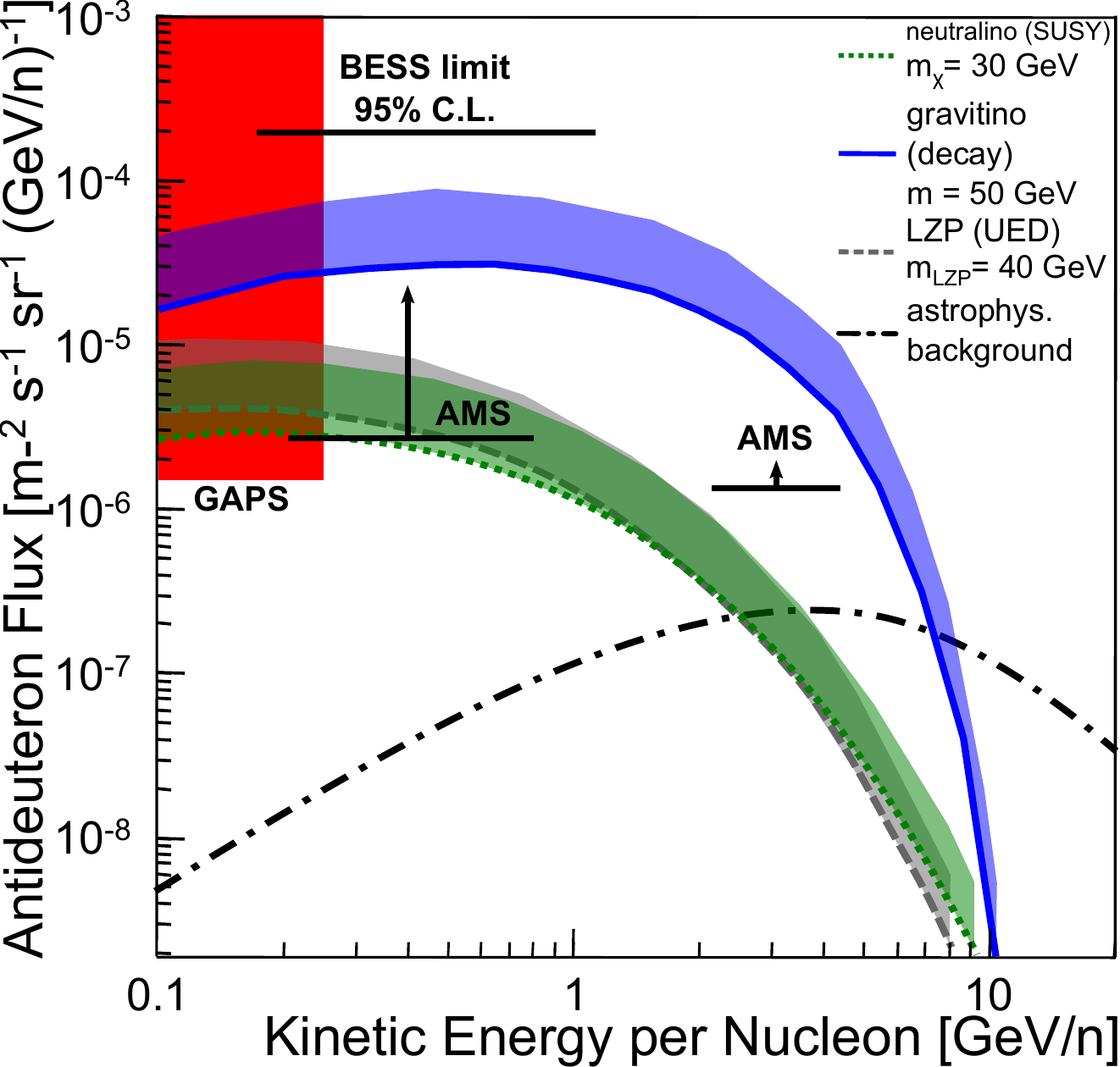}
\vspace{-0.2in}
\caption{\label{f-dmdbar} An array of representative DM models~\cite{Baer:2005tw,Donato:2008yx,Dal:2014nda,Ibarra2013a} is compared to the projected sensitivities ($3\sigma$ discovery limit) of three 35-day GAPS flights and five years of AMS-02 operation (\cite{ams02dbaricrc2007}, updated for measurement time, geomagnetic cutoff model, background fluxes, with the arrow showing the AMS-02 geomagnetic cutoff correction~\cite{Aramaki2015}). Also shown are limits from BESS~ \cite{Fuke:2005it}. The shaded regions indicate the theoretical Galactic propagation uncertainties, spanning the conservative MED and optimistic MAX models~\cite{Donato:2003xg}.}
\end{wrapfigure}

The coming decade is an exciting time for experimental searches for cosmic antinuclei, with experiments beginning to be sensitive to viable DM models. The multi-purpose experiment AMS-02 is accumulating large statistics on the International Space Station, and the GAPS Antarctic balloon mission, the first experiment optimized specifically for low-energy ($E<0.25$\,GeV/n) cosmic antinuclei, is preparing for its initial flight in late 2020 (see Fig.~\ref{f-dmdbar}).
To probe this unique parameter space more deeply, the lifetimes of these current missions must be supported and future mission technologies and concepts must be developed in parallel.

It is crucial to utilize many experimental techniques and to explore as many cosmic signatures as possible to fully cover the DM phase space. In recent years, there have been tantalizing hints of DM detection, for example in cosmic-ray positron~\cite{ams02first,ams4,Abeysekara:2017old}, gamma-ray~\cite{Hooper:2011ti,Daylan:2014rsa,2015PhRvL.114u1303K,2015arXiv150205703A,TheFermi-LAT:2015kwa,Calore:2014nla,TheFermi-LAT:2017vmf}, and antiproton~\cite{2015arXiv150405937H,Chen:2015cqa,PhysRevLett.118.191101,PhysRevLett.118.191102,Cuoco:2019kuu,Cholis:2019ejx} spectra, but all have relied on precise modeling of uncertain astrophysical backgrounds. Antideuteron searches, which have exceptionally suppressed astrophysical background, do not suffer such difficulties~\cite{Donato:1999gy,Baer:2005tw,Donato:2008yx,Duperray:2005si,Ibarra2013a,Fornengo:2013osa}. Instead of a small deviation on top of large background spectrum, DM annihilation or decay can produce an antideuteron signal that is orders of magnitude above the background. In addition, the recent announcement by AMS-02 of candidate events with mass and charge consistent with antihelium~\cite{antihe,antihe2} further motivate independent antideuteron and antihelium searches. Though these candidates are tentative, requiring verification or refutation with a complementary experimental technique such as GAPS, a positive signal would refashion the field of cosmic-ray physics and potentially revolutionize the understanding of Big Bang nucleosynthesis~\cite{2018arXiv180808961P}.

\vspace{-0.1in}
\subsection{New Discovery Space with Antideuterons and Antiprotons\label{s-dbar}}

The unique strength of a search for low-energy antideuterons lies in their ultra-low astrophysical background~\cite{Donato:1999gy,Fornengo:2013osa,Baer:2005tw,Donato:2008yx,Duperray:2005si,Ibarra:2012cc}. Secondary/tertiary (background) antideuterons are produced when cosmic-ray protons or antiprotons interact with the interstellar medium~\cite{Duperray:2005si}.
However, the high energy threshold for antideuteron production and the steep energy spectrum of cosmic rays mean that there are few particles with sufficient energy to produce secondary antideuterons, and those that are produced have relatively large kinetic energy.
Fig.~\ref{f-dmdbar} illustrates the importance of low-energy antideuteron searches: the fluxes from a range of viable DM models~\cite{Baer:2005tw,Donato:2008yx,Dal:2014nda} exceed the astrophysical background~\cite{Ibarra2013a} by more than two orders of magnitude in the energy range below a few GeV/$n$. This is in strong contrast to positrons, antiprotons, and $\upgamma$-rays, where only a small contribution on top of the background is expected in optimistic scenarios.
Cosmic-ray antideuterons from any source have never yet been detected, with the best current limits given by BESS~\cite{Fuke:2005it}. 

The generic sensitivity of such a search is illustrated using three specific DM models in Fig.~\ref{f-dmdbar}.  These include a lightest supersymmetric particle (LSP) neutralino~\cite{Donato:2008yx}, with mass and annihilation cross section consistent with the long-standing observed excess of gamma-rays seen by Fermi in the Galactic center~\cite{Hooper:2011ti,TheFermi-LAT:2017vmf}. This excess has been interpreted as originating from DM with mass \texttildelow25--45\,GeV~\cite{Hooper:2011ti,Daylan:2014rsa}. Observations of the Andromeda galaxy have revealed a similar excess~\cite{Ackermann:2017nya,Karwin:2018lww}, giving further motivation to this possible DM detection. While it is notoriously difficult to eliminate other possible sources of gamma-ray emission, such as pulsars~\cite{2015arXiv150402477O,Lee:2015fea,Bartels:2015aea,Haggard:2017lyq}, antideuteron searches can either confirm or rule out the DM origin of this signal. Also shown are a right-handed Kaluza-Klein neutrino of extra-dimensional grand unified theories (LZP)~\cite{Baer:2005tw}, which is an example of non-supersymmetric DM, and a decaying LSP gravitino~\cite{Dal:2014nda}, which cannot be seen by direct detection experiments. Direct searches are most sensitive to certain couplings of DM to nuclei, but are mostly insensitive to others; thus antideuteron searches can identify DM models, such as this decaying gravitino, that are otherwise inaccessible.

\begin{figure}
\centering
\includegraphics[width=0.8\linewidth]{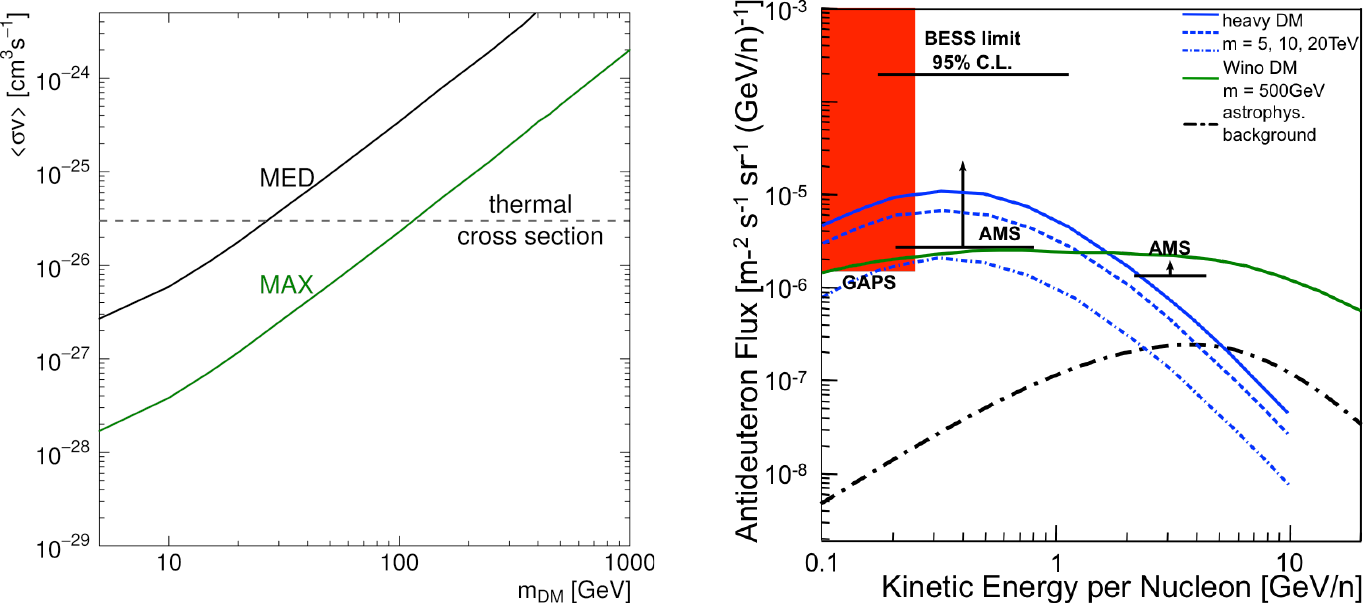}
\caption{\label{f-lightheavy} The sensitivity of the GAPS antideuteron search to light WIMPs annihilating into $u\bar{u}$ for the MED and MAX Galactic propagation scenarios~\cite{2015arXiv150507785A} (left), and the sensitivity of GAPS and AMS-02 to TeV-scale WIMPs annihilating into $b\bar{b}$~\cite{Ibarra:2009tn} and to 500~GeV pure-Wino dark matter~\cite{Hryczuk:2014hpa}, both with Sommerfeld-enhanced annihilation cross-sections in the MAX propagation scenario. }
\vspace{-0.2in}
\end{figure}

Additional models, which range in mass from a few~GeV to several~TeV (see Fig.~\ref{f-lightheavy}), further illustrate the broad range of DM candidates that antideuteron searches can probe.
The experimental sensitivity reaches well below the thermal relic cross section for annihilation into light quarks for DM with mass below 20--100~GeV, depending on the cosmic-ray propagation scenario.
A detectable antideuteron signal may also result from annihilation in certain heavy (5--20\,TeV) WIMP models, if one assumes a cosmic-ray propagation scenario with a larger halo size, as favored by AMS-02 antiproton data~\cite{Brauninger:2009pe}. In addition, models that combine TeV-mass, pure-Wino DM with Sommerfeld enhancement mechanisms can result in accessible antideuteron signals~\cite{Hryczuk:2014hpa}. These models are motivated by the DM interpretation of the excess of positrons measured by PAMELA~\cite{pamela}, Fermi~\cite{fermi}, and AMS-02~\cite{ams02first,ams02elposflux2014,ams4}, which has been difficult to resolve due to possible positron background from pulsars. Heavy DM models are particularly challenging for direct detection and collider methods. 
Antideuterons also provide a unique avenue to explore candidates in the next-to-minimal supersymmetric model (NMSSM), which evade direct and collider bounds~\cite{Cerdeno:2014cda}. 
Antideuteron searches will thus probe models of new physics over a wide mass range that have proven difficult to resolve with other search methods. 

The theoretical uncertainties of antideuteron formation and cosmic-ray propagation are both $\mathcal{O}(10)$~ \cite{2015arXiv150507785A}.  The propagation uncertainty is indicated by the shaded regions of Fig.~\ref{f-dmdbar}, which span the difference between using the conservative MED and the optimistic MAX model to describe the propagation of charged particles to Earth from where they are produced in the Galactic halo~\cite{Donato:2003xg}. 
Recent positron observations exclude the MIN Galactic propagation model~\cite{2012JCAP...01..049B,2014PhRvD..90h1301L}, and recent AMS-02 antiproton results have been interpreted to favor the MAX model~\cite{2015arXiv150404276G}. 
The formation of light (anti)nuclei in hadronic interactions is described by different models (coalescence or thermal) and is not well understood~\cite{diego}.
Despite these large theoretical uncertainties, the upcoming decade of antideuteron searches will be able to definitively probe many DM scenarios.

Low-energy antiproton measurements are an essential partner to these antideuteron searches, as any antideuteron detection must be consistent with antiproton search results.  
A precision low-energy antiproton measurement will also be essential to understand the propagation of all charged particles in the Galactic and Solar environments. In addition, since the antiproton spectrum from DM annihilations shifts towards lower energies with decreasing DM mass, precision measurements of the low-energy antiproton spectrum offer new phase space for probing light DM models~\cite{Aramaki:2014oda} and primordial black holes~\cite{Page:1976wx,Maki:1995pa,2002A&A...388..676B}.
The PAMELA\cite{2010PhRvL.105l1101A,Adriani:2012paa}, BESS\cite{besspbar}, and AMS-02\cite{PhysRevLett.117.091103} antiproton results have all been used to constrain DM models, as well as astrophysical production and propagation scenarios~\cite{Fornengo:2013osa,pbarci,Hooper:2014ysa,Boudaud:2014qra,Cerdeno:2014cda}. 
In addition, recent studies of the AMS-02 low-energy antiproton spectrum indicate a slight excess consistent with $\sim$20--80\,GeV DM~\cite{PhysRevLett.118.191101,PhysRevLett.118.191102}, invoking some of the same models indicated by the gamma-ray excess from the Galactic center.
The high-statistics antiproton spectrum provided by GAPS, in an energy range below that of any other experiment, will be able to confirm or exclude this low-energy antiproton excess. 

\vspace{-0.1in}
\subsection{Cosmic-Ray Antihelium?}

Antihelium from antimatter-dominated regions of the universe is nearly excluded~\cite{Adams:1997ym,Cohen:1997ac}, but recent work shows that antihelium could form from DM annihilation~\cite{Cirelli:2014qia, Carlson:2014ssa, Coogan:2017pwt}.  Recently, AMS-02 has announced the observations of ``a few" candidate events consistent with cosmic antihelium~\cite{antihe,antihe2}. Although these results are not yet published, they have prompted significant theoretical work on the implications for DM models and predicted antideuteron and antiproton fluxes~\cite{Blum:2017qnn, Tomassetti:2017qjk, Lin:2018avl, Korsmeier:2017xzj, 2018arXiv180808961P}. In the coming decade, confirmation or exclusion of this signal using the increased statistics of AMS-02 and the complementary detection technique of GAPS will be essential.
The confirmation of any $Z=-2$ cosmic-ray particle would be transformative, and certainly further motivate the search for antideuterons.

\vspace{-0.1in}
\section{Experiments for Detection of Cosmic-ray Antinuclei}

\vspace{-0.1in}
\subsection{The AMS-02 Experiment\label{s-ams}}

AMS-02 is a multi-purpose cosmic-ray detector that has been operating on the International Space Station (ISS) since May 2011~\cite{ams4}. It is planned to operate AMS-02 until the end of the lifetime of the ISS (at least 2024). In contrast to the high-statistics spectral measurements of other cosmic-ray species, the antideuteron and antihelium studies of AMS-02 are focused on a first-time discovery. AMS-02 follows the principle of typical magnetic spectrometer particle physics detectors, with particle identification that relies on combining signals from an array of sub-detectors.

The transition radiation detector is used to suppress particles of lower mass (like electrons, pions, and kaons). It is also used to identify hard interactions, which could lead to production of antinuclei inside the instrument. The time-of-flight (TOF) system provides the main trigger for the experiment and determines the velocity of the particle up to $\beta\approx0.8$. The particle momentum can be extracted from its trajectory in the magnetic field (\texttildelow0.15\,T). In the low-momentum region multiple scattering becomes an important effect that limits the mass resolution. In the high-velocity region two different types of ring imaging Cherenkov (RICH) counters are used (NaF, aerogel) for the velocity measurement. 
Depending on the underlying propagation and antinuclei formation models, the RICH range can be capable of probing the astrophysical antinuclei production.

\vspace{-0.1in}
\subsection{The GAPS Experiment\label{s-gaps}}

The General Antiparticle Spectrometer (GAPS)~\cite{Aramaki2015} is optimized specifically for low-energy ($<0.25$~GeV/n) cosmic-ray antinuclei. The experiment consists of ten planes of semiconducting Si(Li) strip detectors surrounded on all sides by a plastic scintillator TOF. GAPS is scheduled for its first Antarctic balloon flight from McMurdo in late 2020/early 2021.

GAPS relies on a novel particle identification technique based on exotic atom formation and decay~\cite{Aramaki2015}. First, a low-energy antiparticle that has been slowed by the atmosphere passes through the TOF system, which provides a high-speed trigger and measures particle velocity and $\text dE/\text dx$. It is further slowed by $\text dE/\text dx$ losses in the Si(Li) detectors, eventually stopping inside the detector. It then, with near unity probability, replaces a Si shell electron to form an exotic atom in an excited state. This exotic atom then de-excites through auto-ionization and radiative transitions, emitting X-rays. These X-ray energies are uniquely determined by the antiparticle and Si reduced mass and atomic numbers. Finally, the antiparticle annihilates with the Si nucleus, producing a nuclear star of pions and protons.
The simultaneous occurrence in a narrow time window of X-rays of characteristic energy and nuclear annihilation products with measured multiplicity provides an enormously constraining signature to distinguish antiparticles and to suppress non-antiparticle background. The main challenge for identifying antideuterons is the rejection of the dominant antiproton background, but the combination of exotic atom signatures provides the required rejection.

This exotic atom detector design yields a large grasp compared to typical magnetic spectrometers, and allows for the identification of antiproton, antideuteron, and antihelium cosmic rays. GAPS will provide a precision antiproton spectrum in the low-energy range below $0.25$~GeV/$n$.  Due to the higher charge, background from antiprotons is less of a concern for the antihelium analysis, which allows for a competitive antihelium sensitivity in the low-velocity range ($\beta<0.5$). 

\vspace{-0.1in}
\subsection{Complementary Technologies for Rare Antinuclei Searches\label{s-comp}} 

Any rare event search requires multiple experiments to confirm detections or exclusions of potential signals. Separate experimental designs yield different backgrounds and methods of suppressing these backgrounds, allowing for independent cross-checks of results. 

Suppressing cosmic-ray antiprotons is the biggest challenge for both the GAPS and AMS-02 antideuteron searches. The isotopic separation of (anti)protons from (anti)deuterons is among the most challenging analyses for AMS-02 because the mass resolution is driven by the strength of the magnetic field. Furthermore, a magnetic spectrometer must also provide rejection power to distinguish antideuterons from the more than $10^{10}$-times more abundant protons. In GAPS, low-energy protons cannot mimic the exotic-atom signature. The comparison of low-energy (anti)proton fluxes as well as antiproton-to-proton ratios measured by GAPS and AMS-02 with different detection techniques will serve as a critical benchmark to reduce systematic uncertainties for antinuclei searches for both experiments.

AMS-02 and GAPS also have different advantages in terms of the large exposures necessary for rare signal searches. Although the placement of AMS-02 on the ISS allows for long exposure times, the orbit is not favorable for observing low-energy cosmic rays because it passes through a region with a high geomagnetic cutoff. In the AMS-02 low-energy range, only \texttildelow10\% of charged particles reach the detector, which must be corrected by geomagnetic modeling, and, hence, introduces additional systematic effects. In the higher velocity range, AMS-02 provides sensitivity to antinuclei up to about $\beta=0.99$. Although the GAPS exposure time will be shorter than that of AMS-02, the GAPS geometric acceptance is larger and its Antarctic flight path is tailored to low-energy particles, meaning that GAPS will effectively not require geomagnetic cutoff correction.

\vspace{-0.1in}
\section{The Coming Decade and Beyond}

Within the AMS-02 program, antinuclei searches are among the measurements that will benefit the most from extended data taking on the ISS. The ultimate goal of the GAPS program is three 35-day long duration balloon (LDB) flights, beginning at the cusp of this coming decade, providing an antideuteron sensitivity of $1.5\cdot10^{-6}$\,m$^2$sr s GeV/$n$ ($CL>99$\%).

The current experiments sensitive to cosmic-ray antinuclei can be seen as discovery experiments. Successor experiments will need to aim at measuring spectra with high statistics. This is only possible with extended data taking and larger payloads. Therefore, the development of reliable ultra-long duration balloon platforms with low-geomagnetic cutoff trajectories will directly benefit the study of cosmic-ray antinuclei. At the same time, these flights serve as the technical proving ground for large-scale long-duration experiments for the measurement of cosmic rays, e.g., on the Moon's surface or at one of the Lagrange points. The design of such projects will also be informed by the results of the ground-based experiments.

\pagebreak
\bibliographystyle{nature}

\bibliography{bib}

\end{document}